\documentclass[twocolumn,pre,amsmath,amssymb]{revtex4-1}
\usepackage{graphicx}
\usepackage{dcolumn}
\usepackage{bm}

\newcommand{\R}{\textbf{r}}
\newcommand{\norm}{\hat{\textbf{N}}}

\begin{document}

\title{Minimal resonances in annular non-Euclidean strips}
\author{Bryan Gin-ge Chen}
\affiliation{
Department of Physics and Astronomy, University of Pennsylvania, Philadelphia, PA 19104-6396}
\author{Christian D. Santangelo}
\affiliation{
Department of Physics, University of Massachusetts, Amherst, MA  01003}
\date{\today}

\begin{abstract}
Differential growth processes play a prominent role in shaping leaves
and biological tissues. Using both analytical and numerical
calculations, we consider the shapes of closed, elastic strips which
have been subjected to an inhomogeneous pattern of swelling. The
stretching and bending energies of a closed strip are frustrated by
compatibility constraints between the curvatures and metric of the
strip. To analyze this frustration, we study the class of ``conical''
closed strips with a prescribed metric tensor on their center
line. The resulting strip shapes can be classified according to their
number of wrinkles and the prescribed pattern of swelling. We use this
class of strips as a variational ansatz to obtain the minimal energy
shapes of closed strips and find excellent agreement with the results
of a numerical bead-spring model. We derive and test a surprising 
resonance condition for strips with minimal bending energy along 
the strip center line to exist.
\end{abstract}

\pacs{PACS}

\maketitle

The elastic buckling of thin sheets plays an important role in the
shaping of biological tissues, especially the leaves of many plants
\cite{drasdo, mahadevan09,marder03a, sharon07, nath03}. In this
context, the buckling results from the addition of extra material in
the sheet, as when the sheet is subjected to local growth
\cite{benamar05}. Recent experiments have also demonstrated that such
growth processes can be mimicked by synthetic systems \cite{swinney02,
mora06, klein07}.  There have been several different theoretical
formulations used to predict shapes of buckled objects which, broadly,
fall into two classes. The first approach is to incorporate swelling
by modifying the F\"oppl-von K\`arm\`an equations
\cite{landau,mahadevan09,benamar05,benamar08}. This is especially
suited to studying the stability of nearly flat sheets but its
applicability for studying the post-buckling behavior of swelled
sheets is not obvious. A second approach defines a metric, prescribed
by the local swelling of the sheet, from which an ``ideal,''
strain-free shape can be determined \cite{nechaev,audoly02, audoly03,
marder03, marder06, efrati08a, santangelo09}. This approach is best
suited for thin sheets, where one can expect the sheets to be nearly
strain-free. 

In this paper, we will continue an approach, begun in Ref.
\cite{santangelo09}, to study narrow strips that have been subjected
to an inhomogeneous pattern of swelling. We will focus on the specific
case of closed, unknotted strips such as one that may be cut from the
edge of a circular disk. 
Based on the results of a direct numerical
minimization of a model of a swelled strip, we will focus on a class
of strips for which the normal vector on the center line always lies
tangent to a fixed cone: a ``conical'' strip. Within this class of
conical strips, we are able to find profiles, depending only on the
metric tensor along the center line, that agree favorably with
numerical minimization of a bead-spring model.
This analysis extends the results of Ref. \cite{santangelo09} to the entire
class of conical strips, and corroborates those results numerically.

For very thin, open strips, it can be shown that near-isometric embeddings
exist both with vanishing mean curvature and mean curvature gradients
along the center line \cite{kupferman}. Nevertheless, we derive a ``resonance''
condition, written in terms of the metric and the number of wrinkles
in the embedding, for which a {\em closed}
conical strip can have vanishing mean curvature along its center line.  Our
results indicate that the additional
topological condition that the strip be closed results in nonvanishing
mean curvature -- thus, topology frustrates the bending energy.  This
prediction is in agreement with our numerical results.

Throughout this paper, we will specialize to the case that the
Gaussian curvature is negative. As a motivational example, consider
the experiment described in Klein \textit{et al} \cite{klein07}, in
which a thin film of NIPA polymer gel is endowed with a controlled,
axisymmetric monomer concentration that varies as a function of
distance from the disk center. When the temperature is increased, the
disks shrink to a configuration in which $K<0$ is approximately constant.

It is
fortuitous that there are families of smooth surfaces that realize shapes of
constant $K<0$ \cite{embeddingbook, poznyak, arizona}, and, na\"ively, one might
expect a sufficiently thin sheet to adopt one of these known shapes.
The reality, however, is that the experimental sheets seem to adopt
extremely wrinkled morphologies that are quite different from the known smooth embeddings.
Though our results are not directly applicable to these experiments because they are for narrow strips,
they do shed some light on the nature of the frustration. In
particular, we find numerically that the strips allow localized
stretching in order to lower their bending energy, despite the ready
existence of embeddings that admit no stretching at all. This detail will be the subject of a future publication \cite{chen}.

In section \ref{sec:preliminaries}, we frame the problem of inhomogeneous swelling in terms of geometry and isometric embeddings.  In section \ref{sec:closedstrips}, we specialize to closed strips
with inhomogeneous patterns of swelling and introduce the class of ``conical strips''.
In section \ref{sec:numerics}, we describe our direct numerical
minimizations and compare with analytical predictions.  Finally, in
section \ref{sec:conclusion}, we discuss our results.

\section{Swelling elastic sheets}\label{sec:preliminaries}

We will first consider the general problem of determining the shapes
of elastic sheets with inhomogeneous patterns of swelling. We review
the elasticity of such swelled sheets in terms of a
``preferred'' metric assigned to the sheet, and discuss the obstructions to minimizing both the stretching and bending energies. Finally, we derive a simple model to describe the shape of the center line of a non-Euclidean strip.

\subsection{Elasticity as differential geometry}

\begin{figure}
\includegraphics[width=2.5in]{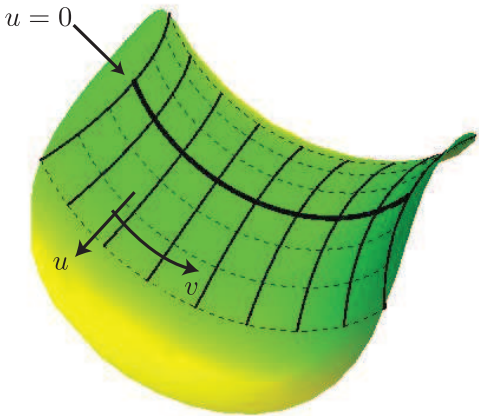}
\caption{\label{fig:coords} (color online) Geodesic coordinates can be constructing starting from a curve (thick line) parametrized by $v$, at which $u=0$. The geodesics perpendicular to the $u=0$ curve are parameterized by their arc length $u$ (thin lines). The curves perpendicular to the geodesics (dashed lines) constitute a coordinate system in which the metric takes the form of Eq. (\ref{eq:geodesiccoords}).}
\end{figure}

In this section, we review the elasticity of swelled sheets, and
establish notation for the rest of the paper. Consider an elastic
sheet with thickness $t$ that has swelled by the position-dependent scalar $\Omega$. For thin sheets, the energy can be decomposed into two terms, $E = E_C + E_B$, the in-plane stretching energy $E_C \propto t$, and the bending energy $E_B \propto t^3$ \cite{landau,efrati08a,ciarlet}. This swelling factor measures the change of infinitesimal distances on
the sheet, and can be converted to a ``prescribed'' metric tensor,
$\bar{g}_{i j}$. When the actual metric (tensor) $g_{i j}$ equals the
prescribed metric $\bar{g}_{i j}$, the in-plane stretching energy,
$E_C$ vanishes. Therefore, $E_C$ is written in terms of a strain
tensor, $\gamma_{i j} \equiv g_{i j} - \bar{g}_{i j}$. When
$\Omega=1$, $\bar{g}_{i j} = \delta_{i j}$ in an appropriate
coordinate system, and this recovers the standard formula for the strain of an elastic sheet \cite{ciarlet}.

It will prove convenient to consider a coordinate system $(u,v)$ in which the metric of the buckled surface takes the form
\begin{equation}\label{eq:geodesiccoords}
g_{u u} du^2 + 2 g_{u v} du dv+g_{v v} dv^2 = du^2 + \rho^2(u,v) dv^2.
\end{equation}
This can be compared to the metric on a flat disk
$du^2+u^2dv^2$, where $u,v $ are the radius and polar angle in polar 
coordinates, respectively.  

Coordinate systems of these types always exist locally and can be
found by the construction illustrated
in Fig. \ref{fig:coords} (also see page 286 of Ref. \cite{DoCarmo-book}).  
First, choose an arbitrary curve on the surface to lie along $u=0$,
and let $v$ be the parameter along that curve. For instance, this curve will be the center line of the strips
that we consider.  Curves of
constant $v$ are defined to be geodesics perpendicular to the $u=0$
curve, parametrized by their arc length $u$. Curves of
constant $u \ne 0$ are thus curves of fixed distance $u$ from the
$u=0$ curve along that family of geodesics. Since the resulting curves are mutually
orthogonal and $u$ is arc length, the metric must have the form of Eq.
(\ref{eq:geodesiccoords}). By Gauss's \textit{theorema egregium}, the
Gaussian curvature can be written in terms of the metric alone as $K =
-\partial_u^2 \rho/\rho$. The function $\rho(0,v)$ is determined by
the metric on the curve at $u=0$. We will always choose the
coordinate $v$ to be dimensionless, so that $\rho$ carries dimensions
of length.
Here, $\rho(u,v) dv$ is the infinitesimal length along curves of constant $u$ and the
width, $w$, is the width of the strip \textit{after} swelling. 

To quadratic order in the strain, the elastic energy is \cite{efrati08a}
\begin{eqnarray}\label{eq:compressionenergy}
E_C &=& \frac{t Y}{2 (1-\nu^2)} \int du dv \sqrt{g}~\left[\nu \left(\bar{g}^{i j} \gamma_{i j}\right)^2\right.\\
& & \left. +  (1-\nu) \bar{g}^{i k} \bar{g}^{j l} \gamma_{i j} \gamma_{k l} \right],\nonumber
\end{eqnarray}
where $Y$ is the Young's modulus and $\nu$ the Poisson ratio of the sheet. 
The bending energy is given by the usual expression
\begin{equation}\label{eq:bending1}
E_B = t^3 \frac{\kappa}{2} \int du dv \sqrt{g}~\left[4 H^2 - (1-\nu) K \right],
\end{equation}
where $\kappa \equiv Y/[12 (1-\nu^2)]$, $H = g^{i j} h_{i j}/2$ is the
mean curvature and $K$ is the Gaussian curvature. These are both
written in terms of the second fundamental form, $h_{i j} = \norm
\cdot \partial_i \partial_j \mathbf{r}$ where $\norm$ is the unit surface normal.
Couplings between $\gamma_{i j}$ and $h_{i j}$ cannot exist if the
swelling is invariant under a change in surface orientation, $\norm
\rightarrow -\norm$. We will also neglect variations of the sheet thickness with swelling.

The total energy $E=E_C+E_B$, or some variant of it, has been
considered by a number of authors \cite{sharon07, nechaev,
marder03,marder06,audoly03}. This functional provides a natural generalization of the usual elastic energy to non-Euclidean metrics.

\subsection{Geometrical constraints}
\label{subsec:constraints}

Minimizing this energy is difficult because the
metric and curvature tensors are not independent when a surface is
immersed in three dimensions. In fact, they satisfy a set of
compatibility relations, which we will briefly review in this section.
To begin, we write two equations for how the frame $(\partial_u
\textbf{r}, \partial_v \textbf{r}/\rho, \norm)$ changes \cite{weingarten},
\begin{eqnarray}\label{eq:evolutionu}
\partial_u \left(
\begin{tabular}{c}
$\partial_u \R$\\
$\partial_v \R/\rho$\\
$\norm$
\end{tabular}
\right) = \textbf{R}_u \left(
\begin{tabular}{c}
$\partial_u \R$\\
$\partial_v \R/\rho$\\
$\norm$
\end{tabular}
\right),
\end{eqnarray}
and
\begin{eqnarray}\label{eq:evolutionv}
\partial_v \left(
\begin{tabular}{c}
$\partial_u \R$\\
$\partial_v \R/\rho$\\
$\norm$
\end{tabular}
\right) = \textbf{R}_v \left(
\begin{tabular}{c}
$\partial_u \R$\\
$\partial_v \R/\rho$\\
$\norm$
\end{tabular}
\right),
\end{eqnarray}
where
\begin{eqnarray}
\textbf{R}_u = \left(
\begin{tabular}{ccc}
$0$ & $0$ & $h_{uu}$\\
$0$ & $0$ & $h_{u v}/\rho$\\
$-h_{u u}$ & $-h_{u v}/\rho$ & $0$
\end{tabular}
\right),\label{eq:A}
\end{eqnarray}
and
\begin{eqnarray}
\textbf{R}_v = \left(
\begin{tabular}{ccc}
$0$ & $\partial_u \rho$ & $h_{v u}$\\
$-\partial_u \rho$ & $0$ & $h_{v v}/\rho$\\
$-h_{v u}$ & $-h_{v v}/\rho$ & $0$
\end{tabular}
\right).\label{eq:B}
\end{eqnarray}
These follow from the definitions of the metric and second
fundamental forms in our coordinates.  

These two matrix equations are only compatible if $\partial_u \mathbf{R}_v = \partial_v \mathbf{R}_u + [\mathbf{R}_u,\mathbf{R}_v]$, where $[,]$ is the commutator. This results in three equations relating the second fundamental form to the metric, the Gauss-Codazzi-Mainardi-Peterson equations \cite{weingarten},
\begin{eqnarray}
\frac{1}{\rho} \partial_u \left(\rho h_{u v}\right) &=& \partial_v h_{u u}\nonumber\\\label{eq:gauss-codazzi}
\partial_u \left( \frac{h_{v v}}{\rho} \right) &=& h_{u u} \partial_u \rho + \partial_v \left(\frac{h_{u v}}{\rho}\right)\\
\rho \partial_u^2 \rho &=& h_{u v}^2 - h_{u u} h_{v v}.\nonumber 
\end{eqnarray}
The last equation is the \textit{theorema egregium} of
Gauss, and will be used to relate the function $\rho$ to the functions
$h_{ij}$. Eqs. (\ref{eq:gauss-codazzi}) and (\ref{eq:evolutionu}) can
also be used as evolution equations to construct isometric embeddings
for arbitrary $\rho$ provided that $h_{v v} \ne 0$ (see, for example,
\cite{marder06} and \cite{santangelo09}). 

Finally, we will find it useful in Sec. \ref{sec:closedstrips} to relate 
the components of the
curvature directly to the mean curvature, $H$, given by $2 H \equiv
g^{i j} h_{i j} = h_{u u} + h_{v v}/\rho^2$. Using this relation in
the \textit{theorema egregium}, we find
\begin{equation}\label{eq:algebraic}
2 \rho H \left( \frac{h_{v v}}{\rho}\right)= h_{u v}^2 - \rho \partial_u^2 \rho + (h_{v v}/\rho)^2.
\end{equation}
Therefore $h_{v v}/\rho$ and $h_{u v}$ lie on a circle and
\begin{eqnarray}
h_{u u} &=& H + \sqrt{H^2 - K} \sin \alpha\nonumber\\ \label{eq:ll}
\frac{h_{v v}}{\rho^2} &=& H - \sqrt{H^2-K} \sin \alpha\\\
\frac{h_{u v}}{\rho} &=& \sqrt{H^2 - K} \cos \alpha.\nonumber
\end{eqnarray}
Here, $\alpha$ tells us how the principal curvature axes are aligned
with respect to the coordinate system. For example, when
$\alpha=\pi/2$, the principal curvature $H + \sqrt{H^2 - K}$ points along the tangent vector in the $u$ direction.

\subsection{Non-Euclidean strip energy}\label{eq:energy}
%

In this paper, we will consider the lowest order terms of the stretching and bending energies in an expansion in powers of the ribbon width. Our procedure can be justified by a careful perturbation expansion of both the stretching and bending energies in powers of the strip width $w$, valid as long as $w \sqrt{-K} \ll 1$, where $K$ is the typical Gaussian curvature associated with the prescribed metric. This expansion also yields corrections to the strain along the center line due to the bending energy and will be explored elsewhere \cite{chen}.

For our purposes, we need the following fact: as the ribbon thickness is decreased, its metric more closely agrees with the prescribed metric \cite{efrati08a, santangelo09}.
Therefore, we will assume our strips are sufficiently thin that the strain along the center line vanishes, implying that the Gaussian curvature is equal to the prescribed curvature. From the numerics in Sec.~\ref{sec:numerics}, the assumption of vanishing strain along the ribbon center line is roughly borne out: as the thickness decreases, the strain along the center of the strip similarly decreases. According to Fig. \ref{fig:Kseries}, however, the Gaussian curvature mismatch, implying strain, remains along the edge. We will see that even the condition that $K$ is isometric on the center line is only approximately true for the closed strips we are able to simulate numerically. Nevertheless, the small remaining strain leads to only small quantitive errors.

The size of this error can be estimated if 
we note that Eqs. (\ref{eq:gauss-codazzi}) can be used to obtain the second fundamental form for a fixed metric. 
For clarity, we will suppress all dependence on $u$ unless otherwise
indicated until section \ref{subsec:examples}, as all quantities in
our analytical model will be evaluated on the center line at $u=0$.
If we suppose that the metric deviates from the prescribed metric at the center line, we can obtain an upper bound on the strain. To do so, we suppose a form for the strain for which $\gamma_{u u} = \gamma_{u v} = 0$. Then $\gamma_{v v} =\rho^2 - \bar{\rho}^2$. Since the Gaussian curvature determines $\partial_u^2 \rho$, $\gamma_{v v} \sim \Delta \gamma~ (u/\bar{\rho})^2$, where $\Delta \gamma$ describes the leading order deformation of the strain. Substituting this into the stretching energy in Eq. (\ref{eq:compressionenergy}), we obtain a rigorous upper bound on the stretching energy
\begin{equation}\label{eq:EC}
E_C \sim Y \frac{t w^5}{\bar{\rho}^5} (\Delta \gamma)^2,
\end{equation}
where we have dropped a very small numerical factor depending on $\nu$. This result suggests that the magnitude of the energy in the strain can be controlled by adjusting the ribbon width or, alternatively, its thickness.

Similarly, when $K = \bar{K}$ on the center line, similar considerations show that $E_C \sim Y t w^7/\bar{\rho}^7 \Delta \gamma^2$ where $\Delta \gamma$ depends on the third derivative of $\rho$ at the center line. From Gauss' \textit{theorem egregium}, this is linked to derivatives of the curvature with respect to $u$.

We will approximate the bending energy by its leading order power of $w$, which should be valid when $w$ is sufficiently small. For the bending energy, this yield
\begin{equation}\label{eq:bend}
E_B \approx \frac{t^3 \kappa w}{2} \int dv~\rho(0,v) H^2(0,v),
\end{equation}
where the factor of $w$ accounts for the integral over $u$.  Thus, using Gauss' \textit{theorem egregium} again, this only depends on $\partial_u^2 \rho$ along the center line.

This suggests that the lowest order terms in the energy are minimized when the curvature along the center line minimizes Eq. (\ref{eq:bend}) subject to the constraint on the Gaussian curvature. The higher-order terms

This type of approximation has also been used to describe narrow
developable strips. For instance, to obtain the Sadowsky
functional \cite{sadowsky}, we would minimize Eq. (\ref{eq:bend}) subject to the constraint that $K=0$ on the center line.
In Sec. \ref{sec:numerics}, we will compare our results to a numerical minimization of several closed strips. This will both justify our assumptions as well as confirm the predictions obtained from our calculations.

For infinitely thin open strips of sufficient narrowness, it is possible to find an isometric immersion such that both $H=0$ and $\partial_u H = 0$ along the center line \cite{kupferman}. Most of our numerical strips do not even have $H=0$ on the center line; in fact, we will derive conditions in Sec.~\ref{sec:closedstrips} under which it is possible to at least ensure that $H=0$ on the center line.

\section{Closed strips}\label{sec:closedstrips}

We consider strips that are topologically closed and unknotted,
such as strips resulting from punching a hole into the center
of a disk or swelling a cylinder. Though we will formulate the problem
quite generally at first, we quickly find ourselves unable to generate
sufficient and necessary conditions for a strip with a prescribed
metric to close. To remedy this, we will restrict consideration to a
large family of strips that yield an analytically tractable way to
enforce both the prescribed Gaussian curvature and the closure constraint. We will therefore generate sufficient, but not necessary, conditions for a strip of this type to close.

These strips will have the property that their surface normal on the
center line always points along a cone. Though
many of the results in this section are
not a priori valid for all closed strips, we will show in Sec.
\ref{sec:numerics} that the numerically minimized strips seem to have
this property as well.  From this assumption, we derive several
predictions.  In particular, a key result of this section is that the strip center line can be minimal (\textit{i.e.} $H=0$) if a ``resonance'' condition
\begin{equation}
m = \frac{\partial_u \rho^2 + \rho \partial_u^2 \rho - 1}{2 (\partial_u \rho \pm 1)},
\end{equation}
relating the number of wrinkles $m$ to the derivatives of the metric
at the center line, is met. We corroborate this somewhat surprising
result and others numerically in the next section.

\subsection{The class of closed, conical strips}

Closed strips cut from an unswelled, flat disk have a natural coordinate 
system $(r,\theta)$, where $\theta$ is the azimuthal angle around the strip.
In the swelled case, it is natural to force the coordinate $v$ to be
$2\pi$-periodic as well. 
Once we determine $\rho$, $\partial_u \rho$, $\partial_u^2 \rho$, and $h_{ij}$ on the center line at $u=0$
that result in a closed strip, Eq. (\ref{eq:evolutionu}), which governs the evolution along lines of constant $v$, will preserve the periodic structure of the strip away from the center line.

We first note that Eq. (\ref{eq:evolutionv}) determining the evolution
of the strip frame along $v$ can be recast as an equation for the
components of the two tangents and the surface normal in space along
the center line.  Taking the $x$-component,
\begin{eqnarray}\label{eq:centerlinetrajectory}
\partial_v \left(
\begin{tabular}{c}
$\partial_u \R \cdot \hat{x}$\\
$\frac{\partial_v \R}{\rho} \cdot \hat{x}$\\
$\norm \cdot \hat{x}$
\end{tabular}
\right) = \textbf{R}_v \left(
\begin{tabular}{c}
$\partial_u \R\cdot \hat{x}$\\
$\frac{\partial_v \R}{\rho} \cdot \hat{x}$\\
$\norm\cdot \hat{x}$
\end{tabular}
\right).
\end{eqnarray}
Similarly, we can replace $\hat{x}$ with $\hat{y}$ or $\hat{z}$. The
skew-symmetric matrix $\textbf{R}_v$ is an infinitesimal rotation that
depends on the second fundamental form along the strip center line,
which itself is a function of $v$. Therefore, Eq.
(\ref{eq:centerlinetrajectory}) can be recast as the equation for a
particle in a magnetic field. To make this mapping explicit, we
identify the $v$ coordinate with time and define a ``time-dependent
magnetic field''
\begin{equation} 
\textbf{B}(v) = -\frac{h_{v v}}{\rho} \hat{i} + h_{u v} \hat{j} -
\partial_u \rho \hat{k},
\end{equation}
and the vector
\begin{equation}
\mathbf{v_x}(v) = \left(\partial_u \textbf{r} \cdot \hat{x}\right)
\hat{i}+\left[\left(\frac{\partial_v \textbf{r}}{\rho}\right) \cdot
\hat{x}\right] \hat{j} + \left(\hat{\mathbf{N}} \cdot \hat{x}\right)
\hat{k}.
\end{equation}
The index on $\mathbf{v_x}$ refers to the fact that it arises from taking
the $x$-component of the frame in Eq. (\ref{eq:centerlinetrajectory}).
The curves $\mathbf{v_y},\mathbf{v_z}$ are defined similarly.
The $\hat{i},\hat{j},\hat{k}$ introduced here label the 
$\partial_u\textbf{r}$, $\frac{\partial_v\textbf{r}}{\rho}$, and $\norm$ 
components in the frame in Eq. (\ref{eq:centerlinetrajectory}) and are not the
directions $\hat{x},\hat{y},\hat{z}$ of our three dimensional
Euclidean space. Eq. 
(\ref{eq:centerlinetrajectory}) now reads
\begin{equation}
\partial_v \mathbf{v_x} = \mathbf{B} \times \mathbf{v_x}.
\end{equation}

So far, the steps we've taken here in no way restrict the generality
of our discussion. Requiring that our surface is a closed strip is equivalent to finding 
conditions on $\textbf{B}$ such that $\mathbf{v_i}(v)$ is a closed curve
and that $\int_0^{2\pi}\rho(v)~ [{\bf v_i}(v)\cdot\hat{j}]dv=0$ for 
$i=x,y,z$.  Supposing this could be done, the
problem would be to minimize the energy with respect to $h_{ij}$ and
$\rho$ on the set of all closed solutions.  However, this is related
to the open Fenchel problem of finding conditions on the curvature and torsion
of a space curve in order for it to close.  Instead of finding all
possible closed solutions, we will construct an ansatz for a class of 
solutions that admit a relatively simple characterization.

We will restrict ourselves to strips whose normal vector, on the center line, is
constrained to lie on a cone.
For convenience, we will refer to these as ``conical strips.''
Though our motivation for considering these solutions was originally
to simplify our analysis, this choice of ansatz turned out to be
particularly apt, as will be shown in Sec. \ref{sec:numerics}.  
In appendix \ref{sec:appendixB}, we derive expressions for conical strips
with general mean curvatures along the center line.  We summarize the
results and notation here.  First, we define
$\sin\xi=H/\sqrt{H^2-K}$ and recall that $H=h_{u u} + h_{v
v}/\rho^2$ and $K=-\partial_u^2 \rho/\rho$ on the center line from
Sec. \ref{subsec:constraints}.  We also have the angle $\alpha$
between the principal curvature axes of the strip and the axes of the $uv$
coordinate system.

The final result for the tangent vector of the center line of the strip is
\begin{eqnarray}\label{eq:tangentline}
\frac{\partial_v \textbf{r}}{\rho} &=& - \sin^2\frac{\phi}{2} \left[\sin (W+\beta) \hat{x} + \cos(W+\beta) \hat{y}\right]\nonumber\\
& & + \cos^2\frac{\phi}{2}\left[ \sin(W-\beta) \hat{x} + \cos(W-\beta) \hat{y} \right]\nonumber\\
& & - \sin\phi \cos \beta\hat{z},
\end{eqnarray}
with a constant cone angle $\phi$ between $\norm$ and $\hat{z}$, satisfying
\begin{align}
\tan\phi&=-\frac{\sqrt{\rho \partial_u^2 \rho}\cos\alpha}{\cos\beta\cos\xi
(\partial_u \rho-\partial_v\beta)},
\end{align}
with $W(v)=\sec\phi\int_0^vdv'[
\partial_{v'}\beta(v')-\partial_u \rho(v')]+W_0$, where $W_0$ is an arbitrary constant,
and $\tan\beta=(\sin\alpha-\sin\xi)/\cos\alpha=h_{vv}/(\rho h_{vu})$.
These expressions can be written directly in terms of the functions
$\rho$ and $h_{ij}$ that characterize the shape of the strip on the
center line.

We now derive conditions under which a conical strip closes. We
first note that the tangent vectors can only be periodic if $\Delta
\beta \equiv [\beta(2 \pi)-\beta(0)]/(2 \pi)$ and $\Delta W \equiv
[W(2 \pi)-W(0)]/(2 \pi)$ are integers. 
Note that if we fix $0<\phi<\pi/2$ then the magnitude of the term with
$\cos^2(\phi/2)$ in Eq. (\ref{eq:tangentline}) is larger, so that the
number of times the strip wraps around will be set by the period of
$W-\beta$ rather than $W+\beta$.  Therefore, we interpret 
$\Delta W-\Delta\beta$ as the number of times the strip wraps around
the $z$-axis. In particular, this implies that if we were to cut a
closed strip from the edge of a disk, we should require $\Delta W -
\Delta \beta = \pm 1$; other choices of $\Delta W - \Delta \beta$
could be used to represent knotted strips. 
In addition, the integer $\Delta \beta$ counts
the number of times that the tangent vector of the center line crosses
the $z=0$ plane, since it does so whenever $\beta$ is an odd integer
multiple of $\pi/2$. Therefore,
we interpret $\Delta \beta$ as the number of wrinkles of a closed strip.
Finally, we define $\Delta \rho \equiv [\int_0^{2 \pi}
dv~\partial_u \rho(v)]/(2 \pi)$. From the definition of $W$,
\begin{align}\label{eq:Wphi}
\Delta W &= \left(\Delta \beta-\Delta \rho \right)\sec\phi\nonumber\\ 
\cos\phi&=\frac{\Delta\beta-\Delta\rho}{\Delta\beta\pm1},
\end{align}
and thus the two choices of $\Delta W$ for a given $\Delta\beta$ mean
that we have two possible cone angles $\phi$ for a given metric that
can lead to closed conical strips.  This gives us a criterion for the
existence of such a strip, since $\Delta \rho$ and
$\Delta \beta$ must be chosen so that $0<\cos\phi<1$ (since
$0<\phi<\pi/2$). In particular, this implies 
\begin{align}
0&<\frac{\Delta\beta-\Delta\rho}{\Delta\beta\pm1}<1
\end{align}
The above conditions ensure that $\partial_v\textbf{r}/\rho$ is
periodic. As a final constraint, $\partial_v\textbf{r}$ must integrate 
to zero as $v$ goes from 0 to $2\pi$ to ensure that $\textbf{r}$ is
periodic as well.

\subsection{Some examples}
\label{subsec:examples}

It is instructive to consider some specific examples to understand the
formulas derived in the last section. Consider first conical strips
with $\Delta \beta = 0$. This class of shapes contains all of the
axisymmetric strips, for which $\beta$ is constant in $v$. With this, 
$0<\Delta \rho < 1$. For axisymmetric strips, for which $\rho$ is
also independent of $v$, this criteria is well-known (see the proof in
Marder \textit{et al.} \cite{physicstoday} for another derivation of this
criteria for axisymmetric strips). 
The class of strips with $\Delta \beta
= 0$ contains, but is not limited to, axisymmetric shapes -- the function
$\beta$ need not be constant; all that is required is that $\beta$ be
periodic as $v$ varies from $0$ to $2 \pi$. The condition $0<\Delta
\rho < 1$ yields a limit on how
quickly $\rho$ can grow with $u$ for such a conical strip.
This has ramifications for a disk with $K = -1/R^2$, which has a
metric with $\rho(r,v) = R \sinh (u/R)$ \cite{klein07, santangelo09}, where $r$
is the radial distance from the disk center. If we imagine cutting a
narrow strip from the edge of the disk with center line at $r=u_0$,
$\partial_u \rho=\cosh(u_0/R)> 1$,
which violates the constraint for $\Delta \beta = 0$.

We next turn our attention to the case of $\Delta \beta = 2$,
corresponding to a saddle shape in which $h_{0,v v}$ has four
zeros (two wrinkles). Then solution in this class of strips must have $1< \Delta \rho<2$ 
or $-1<\Delta\rho<2$.
If we specialize again to a strip cut from a constant negative
Gaussian curvature disk so that $\rho(u,v)=R\sinh((u+u_0)/R)$, 
$\rho = R \sinh(u_0/R)$ and $\Delta\rho=\cosh(u_0/R)$, and we find 
that a solution of the first type is possible when $0< u_0 < R
\cosh^{-1}(1)$ or $0<u_0 <1.32 R$.  At the limiting case,
$\cos\phi=0,1$ and the mean curvature along the center line diverges, as can be readily 
seen from Eq. (\ref{appeq:meancurvature}).

However, this should \textit{not} be interpreted as the maximum radius
of finding a saddle-shaped isometric embedding. In fact, arbitrarily
large isometric embeddings of disks with $K = -1/R^2$ exist \cite{poznyak}. If
we were to cut a strip from the edge of such an isometric embedding
with radius larger than $R \cosh^{-1}(2)$, the strip center line may
still have two wrinkles (alternatively, four zeros for $h_{v v}$).
It could not be a conical strip, however. Consequently, it is likely that 
minimal energy strips deviate from the shapes we would predict for 
conical strips, especially when the mean curvature is large.

We now return to conical strips with $H=0$ along their center line. In
that case, using Eq. (\ref{appeq:meancurvature}), we have 
\begin{equation}\label{eq:closedminimal}
\partial_u \rho - \partial_v \beta = -\cot\phi\sqrt{\rho \partial_u^2 \rho}
\end{equation}
at every $v$. For simplicity, we will specialize to the case that 
$\rho$ is constant in $v$ along the center line, though by
replacing these quantities by their averages over $v$ we could lift
this specialization. In this case, $\Delta \rho = \partial_u \rho$ which implies 
that $\partial_v \beta = \Delta \beta$.
Squaring both sides and using $\cot^2\phi=\frac{\left(\Delta
\beta-\Delta \rho \right)^2}{\left(\Delta \beta\pm1\right)^2 - \left(\Delta
\beta - \Delta \rho\right)^2}$ derived from Eq. (\ref{eq:Wphi}), we find
\begin{equation}
\left[ \left(\Delta \beta\pm1\right)^2 - \left(\Delta \beta - \Delta \rho\right)^2 \right] = \rho \partial_u^2 \rho,
\end{equation}
which implies
\begin{equation}\label{eq:minimalbeta}
\Delta \beta = \frac{(\partial_u \rho)^2 + \rho \partial_u^2 \rho - 1}{2 (\partial_u \rho \pm 1)}.
\end{equation}
Consider again the disk with constant negative gaussian curvature 
$-1/R^2$.  For a strip with center line at radius $u_0$, $\rho = 
R \sinh(u_0/R)$, $\partial_u \rho=\cosh(u_0/R)$, $\partial_u^2 \rho=\rho/R^2$ and the 
resonance condition above is satisfied when $\Delta \beta = 
\cosh (u_0/R) \mp 1$.  This generalizes the condition found by one 
of us numerically \cite{santangelo09}.  For the class of solutions
considered there, this condition could also be derived by Fourier
analysis.

Again, it need not be the case that there are no closed strips with $H=0$ along their center lines
when this condition is not satisfied, only that there are none in the
class of closed, conical strips.  Nevertheless, our results do have
ramifications on the shape of minimal energy strips: we will see in
the next section that we can satisfy both the bending and stretching
energies to lowest order in powers of $w/R$ at one of these
resonances. Therefore, in the regime of thickness we are primarily
concerned with in this paper, $\partial_v \beta = \cosh(u_0/R) \mp 1$ is
\textit{sufficient} to demonstrate that $H=0$ along the center line. 
The numerical evidence we will present in Sec. \ref{sec:numerics}
suggests that, at the very least, embeddings with zero mean curvature
on the center line do not exist for all choices of metric.

\subsection{Energy minimization}\label{sec:ansatz}

We now seek to minimize the bending energy along
the center line of a closed, conical strip.
The energy in Eq. (\ref{eq:bend}) would be minimized when $H=0$ along the center line. Happily, such a condition
can be satisfied within the class of closed, conical strips when Eq. (\ref{eq:minimalbeta}) is satisfied. 
In this section, we are interested in the minimal
energy shape close to $H=0$ as well.

When $H=0$ along the center line, $\beta = m v$ from Eq.
(\ref{eq:closedminimal}). Near an $H=0$ resonance, we can look
for an approximate minimum of the form $\beta = m v + f$, where $\beta
= m v$ is the solution for the $H=0$ strip with $m$ wrinkles. Using
Eq. (\ref{appeq:meancurvature}), we expand $H$ in powers of $f$ to
find
\begin{equation}
H \approx \left[ -\frac{\sqrt{\rho \partial_u^2 \rho}}{\tan\phi\left(\partial_u \rho -
m\right)^2} - \frac{\tan\phi}{\sqrt{\rho \partial_u^2 \rho}} \right] \left[\frac{\partial_v f}{\sin(m v)} \right].
\end{equation}
This allows us to expand Eq. (\ref{eq:bend}) to quadratic order in $f$, from which we find the Euler-Lagrange equation
\begin{equation}
\partial_v \left[\frac{\partial_v f}{\sin^2(m v)}\right]=0,
\end{equation}
which has solution
\begin{equation}\label{eq:perturbcurv}
f = \frac{A}{2} \left[ v + \frac{1}{2 m} \sin(2 m v)\right],
\end{equation}
where $A$ is a constant independent of $v$. Looking at Eq. (\ref{appeq:meancurvature}) for the mean curvature, we see that $\partial_v \beta$ always appears in the combination $\partial_u \rho - \partial_v \beta$. Therefore, we can interpret the first term of Eq. (\ref{eq:perturbcurv}) as a shift in $\partial_u \rho$, which must be accompanied by a $v$-dependent correction proportional to $\sin(2 m v)$. Therefore, when $\partial_u \rho$ is tuned away from a resonance by an amount $A/2$, the function $\beta$ oscillates with a frequency $2 m$.

We can use this near-resonance solution as the basis for a general conical
strip ansatz valid for all metrics: $\beta = m v + \tilde{A} \sin(2 m
v)/(2 m)$. For sufficiently small $\tilde{A}$, this ansatz does not
modify the location of the zeros of $\sin \beta$ nor does it alter
$\partial_v \beta$ at those zeros. It is straightforward to show that
the zeros of $h_{v v}$ coincide with those of $\sin \beta$. At those
zeros, $\partial_v \beta = m - \tilde{A}$. For $H$ to remain finite at
those zeros, inspection of Eq. (\ref{appeq:meancurvature}) implies
that $\partial_v \beta = \partial_u \rho +\cot\phi \sqrt{\rho \partial_u^2 \rho}$ when
$\sin \beta = 0$. Therefore, we conclude that $\tilde{A} = \partial_u \rho
+\cot\phi\sqrt{\rho \partial_u^2 \rho}-m$.

Finally, it is necessary to confirm that this ansatz yields a closed strip. Since $\partial_u \rho$ is independent of $v$, $W = \sec \phi [\partial_u \rho~v - \beta(v) + \beta(0)]+ W_0$. Examining the form of $\beta$, both $W$ and $\beta$ change sign when $v \rightarrow v+\pi/m$. Inserting this into the equation for the tangent vector and integrating, we find that $\int_0^{2 \pi} dv~\rho~\partial_v \mathbf{r}/\rho = \mathbf{r}(2 \pi) - \mathbf{r}(0) = 0$.

\begin{figure}
\includegraphics[width=3.25in]{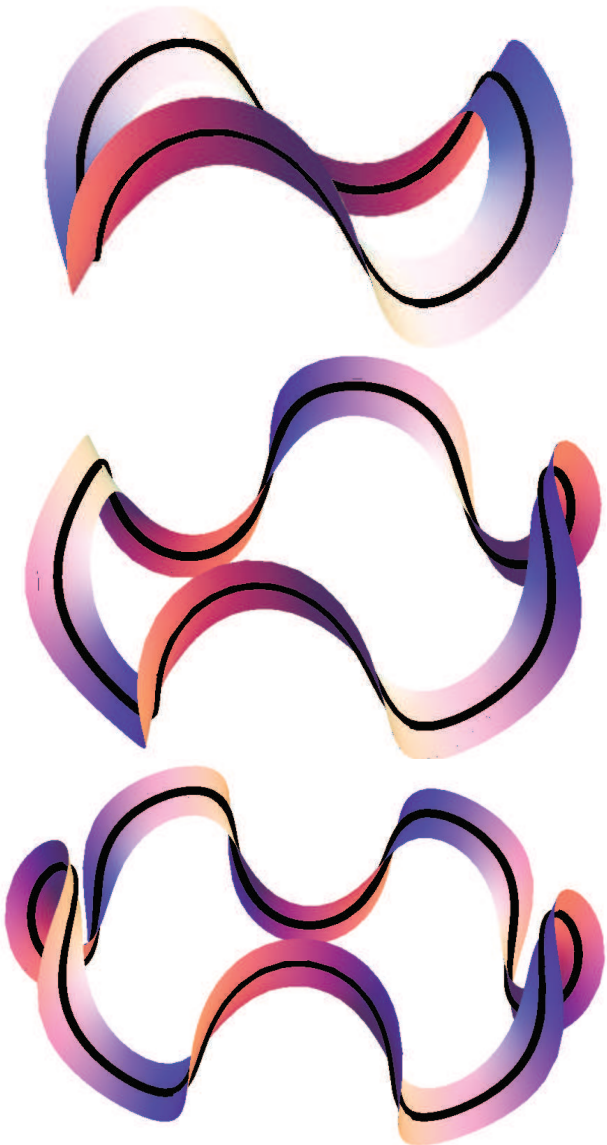}
\caption{\label{fig:m3} (color online) Analytical predictions for 
the shapes of three-, four- and five-wrinkle strips with $H=0$ along
the center line.  The center line is the dark black line, and the
surface displayed is constructed by extending along
the vector $\partial_u{\bf r}$ from the center line.  Here $\rho=R\sinh(u_0/R)$ and $u_0,R$ are chosen to satisfy the
resonance condition $\cosh(u_0/R)=m-1$, when $m$ is the
number of wrinkles.}
\end{figure}

To summarize the construction in this section, assume that we are
given some $\rho(u,v)$.  We would like to construct an ansatz with $m$
wrinkles. 
Next, we calculate $\tan\phi$ using Eq. (\ref{eq:Wphi}).  There
are two choices, which are perhaps not equally good.  Once this choice
is made, it is straightforward to calculate $A$, $\beta$, $W$ and then we
have the expressions for the frame on the strip, including Eq. 
(\ref{eq:tangentline}) for the tangent vector as well as the other 
components, whose expressions are in Appendix
\ref{sec:appendixB}.  If we integrate the tangent vector, we may generate the 
space curve of the center line.  To construct the surface, we integrate Eq.
(\ref{eq:evolutionv}) in a small neighborhood of the center line.  We 
depict center lines of strips with $m=3,4,5$ in Fig. \ref{fig:m3}.  

\section{Numerics}\label{sec:numerics}

\subsection{Numerical model}

To corroborate our analysis, we have implemented the bead-and-spring
model of inhomogeneous swelling developed by Marder and Papaniolaou
\cite{marder06}. An unswelled sheet is represented by a stacked pair of
close-packed triangular lattices with a periodic length of 200 beads
and a width of 20 beads (thus, 8000 particles).  Nearest neighbors in
the lattice are connected by harmonic springs (35000 total bonds) with bond energy between two lattice sites $a$ and
$b$ given by
\begin{equation}\label{eq:bond}
E_{bond} = \frac{1}{2} \left[ \left( \Delta \mathbf{X}\right)^2 - \ell_{ab}^2\right]^2,
\end{equation}
where $|\Delta \mathbf{X}|$ is the bond length. The swelling is incorporated by setting the equilibrium bond length according to the approximate equation
\begin{equation}\label{eq:approximatemetric}
\ell^2 \approx \Delta x^i \Delta x^j \bar{g}_{i j}
\end{equation}
where $\Delta x^i$ are the components of the vector connecting beads $a$ and $b$ in the unswelled membrane and $\bar{g}_{i j}$ is the prescribed metric. This approximates the integral that measures the prescribed distance between points $a$ and $b$. The prescribed metric used in Eq. (\ref{eq:approximatemetric}) is given by the average of the continuum prescribed metric evaluated on the lattice points $a$ and $b$. Alternatively, we could evaluate the continuum prescribed metric somewhere along the bond; as long as the average bond length is small compared to $R$, the details of how we evaluate the prescribed length between neighboring points $a$ and $b$ do not change our results.

As shown in Ref. \cite{marder06}, Eq. (\ref{eq:bond}) is compatible
with a continuum limit using the strain $\gamma_{i j} = g_{i j} -
\bar{g}_{i j}$.
Note that the 2D strip energy consisting of a sum of
stretching and bending energies is recovered from this 3D model when
the thickness is small.  
In particular, we note that the bending energy arises naturally from the differential stretching and compression of the two layers of the sheet.
The energy is minimized using either a conjugate-gradient or
Broyden-Fletcher-Goldfarb-Shanno (BFGS) algorithm, as implemented the
GNU Scientific Library (GSL). In general, we find that strips develop
``creases'' often as the minimization procedure proceeds unless the
strips are sufficiently thick and the change in the degree of swelling sufficiently
small. As a practical matter, therefore, we start our minimization
with thick strips, $t \approx 0.025 w$, and small prescribed
curvatures $\bar{K} \approx - 10^{-2} w^{-2}$. We then increase the
curvature in a sequence of steps to the desired curvature (up to
$\sqrt{\bar{K}}w\approx 1.3$) and, once
this sequence is complete, decrease the thickness to the desired
thickness (down to $t\approx 0.004 w $). If we decrease the thickness first, the strips often
develop creases, though as long as the strips remain smooth during this
process, the energy minima we have found appear to be quite robust
against changes in the sequence of steps taken.
Finally, it is difficult to maintain a smooth strip shape, even at
fixed swelling, once the thickness becomes too small. Therefore, we
are not able to probe the limit of truly thin strips; consequently,
all our strips seem to have some residual strain, even in the ground
state.

We find that strips with different numbers of wrinkles are metastable,
and have explicitly seen this up to five wrinkles. Therefore, to
select the number of wrinkles, we bias the minimization by choosing an
initial strip shape with a prescribed number of wrinkles $m$. Typically,
we use the graph $z(r,\theta) = h_0 \cos (m \theta)$ as an initial
condition for the lattice, where $r$ and $\theta$ are polar coordinates in the $xy$
plane. We have also tried adding additional Fourier modes in the
initial conditions and have
tried amplitudes in the range $0.01 t < h_0 < 100 t$ for
sheets of thickness $t$. In all cases we have tried, the number of
wrinkles is primarily chosen by the dominant mode in $z(r,\theta)$,
though there is a small bias toward lower energy structures with two
wrinkles, and all strips with the same number of wrinkles are nearly
identical.  If no bias is chosen, the minimization will take a long
time to break the symmetry and will go into a state with two or three
wrinkles.

\begin{figure}
\includegraphics[width=3.5in]{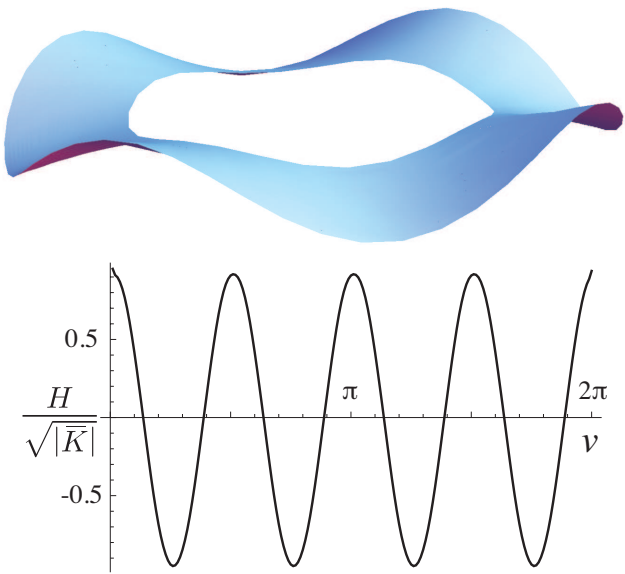}
\caption{\label{fig:threestrip} (color online) A typical minimal
energy strip with four wrinkles and prescribed metric $\rho_1$, with
$\bar{K} = -(1/9) w^{-2}$, $u_0=(5/\pi)w$ and $t=0.01 w$ found by numerical
minimization. Below, the mean curvature of the center line is
displayed.}
\end{figure}

Finally, since we are interested in verifying the assumptions and
predictions of our analytical calculations here, we will focus on
metrics with $\rho(u,v)$ independent of the longitudinal direction $v$. We
have explored two prescribed metrics in detail: (1) $\rho_1(u,v) = R
\sinh(u/R+u_0/R)$ and (2) $\rho_2(u,v) = R \cosh(u/R)/(2 \pi) + \eta R
\sinh(u/R)$. Both of these metrics have $K = - 1/R^2$ everywhere. The
first of these corresponds to cutting a strip of width $w$ from the
edge of a buckled disk at radius $u_0-w/2$. The resonance condition
for the center line of this metric gives $\cosh (u_0/R) = m - 1$. For
$m=3$, this requires a degree of swelling which is difficult to
achieve numerically at low thicknesses. Therefore, to study the resonance condition
numerically, we use $\rho_2(u_0)$ for various values of $\eta$. From
the resonance condition in Eq. (\ref{eq:minimalbeta}), we have four
solutions
\begin{align}
\eta&=
m+(m\pm1)\sqrt{1-\frac{1}{(2\pi)^2(m\pm1)^2}}\\
&\approx 2m\pm1-\frac{1}{2(2\pi)^2(m\pm1)}+\mathcal{O}((m\pm1)^{-2})\\
\text{or }\eta&=
m-(m\pm1)\sqrt{1-\frac{1}{(2\pi)^2(m\pm1)^2}}\label{eq:respred}\\
&\approx \mp1+\frac{1}{2(2\pi)^2(m\pm1)}-\mathcal{O}((m\pm1)^{-2}),
\end{align}
where the $\pm$ must be chosen consistently in each equation.
With the swelling factors accessible to our numerics, we are able to probe 
the branch of resonances with $\eta$ near $+1$ for $2$, $3$, and $4$ wrinkles.

Geometric quantities such as $H$, $K$ for our numerically minimized
surfaces are computed from an interpolation of the points on the mid-surface between 
the two bead-spring sheets.

\subsection{Results}

\begin{figure}
\includegraphics[width=3.25in]{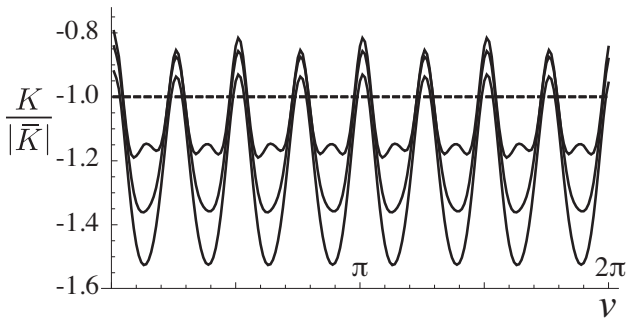}
\caption{\label{fig:4K} Gaussian curvature along the center line for
the strip shown in Fig. \ref{fig:threestrip} with thicknesses $t/w=0.02$, $t/w = 0.015$ and $t/w = 0.01$. As the thickness is lowered, the Gaussian curvature approaches the prescribed curvature (straight dotted line).}
\end{figure}

As an example of a minimial energy surface from our numerical
calculations, Fig.
\ref{fig:threestrip} shows a four wrinkle strip with metric from
$\rho_1$ and $t=0.01 w$ with prescribed Gaussian curvature, $\bar{K}
= - (1/9) w^{-2}$. In Fig. \ref{fig:4K}, we plot the Gaussian
curvature along the center line for strips for three thicknesses
$t/w= 0.02$, $0.015$, and $0.01$ and otherwise identical parameters.
As expexted, along the center line $K$ approaches its prescribed value as the thickness decreases. It is interesting to note that even for the thinnest strips we have tried, the Gaussian curvature always displays localized regions of stretching. Nevertheless, our results are apparently robust against this localized residual strain.

\begin{figure}
\includegraphics[width=3.25in]{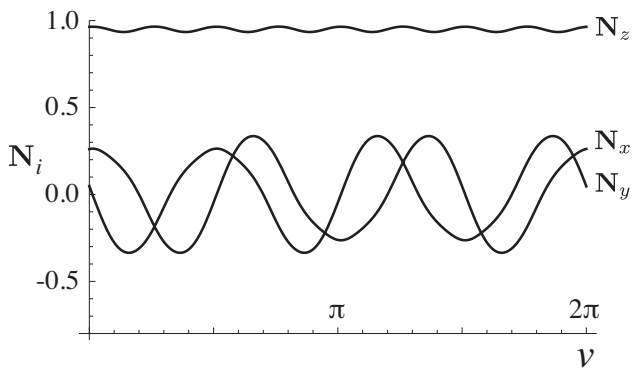}
\caption{\label{fig:norm} The three components of the normal for the
strip of Fig. \ref{fig:threestrip} along the center line, which makes
a nearly constant angle with the $z$-axis. The average $z$-component
is $\langle \mathbf{N} \cdot \hat{z} \rangle = N_z \approx 0.950$
compared to the predicted value $\cos \phi = 0.952$. The variance
$[\langle (\hat{N} \cdot \hat{z} - N_z)^2 \rangle ]^{-1/2} \approx
0.01$.}
\end{figure}

In Fig. \ref{fig:norm}, we plot the three components of the normal
vector on the center line for the thinnest of the strips in Fig. \ref{fig:4K}.
The $x$ and $y$ components oscillate (with amplitude close to $0.25$)
around zero while the $z$ component is very nearly constant. To
quantify this, we compute the average $N_z \equiv \langle \mathbf{N}
\cdot \hat{z} \rangle \approx 0.950$ and the variance $[\langle
(\hat{N} \cdot \hat{z} - N_z)^2 \rangle ]^{-1/2} \approx 0.01$. The
analytical angle $\cos \phi$ is predicted from Eq. (\ref{eq:Wphi}) to be $0.951$. 

\begin{figure}
\includegraphics[width=3.25in]{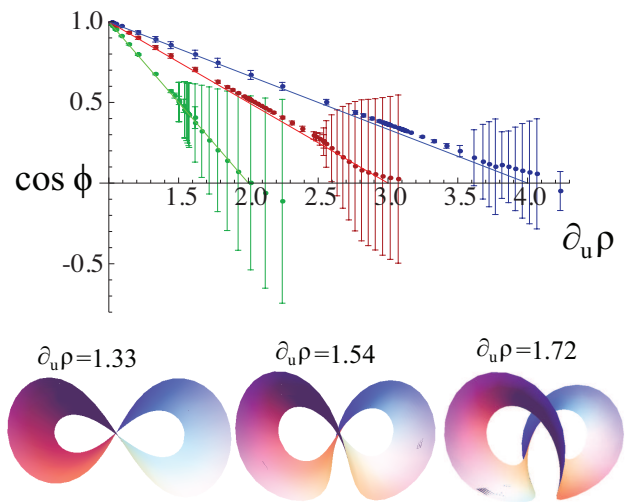}
\caption{\label{fig:anglecompare} (color online) Comparison of the
cosine of the angle of the normal vector to the $z$-axis on the center
line (points) on numerically minimized strips to the prediction of Eq.
(\ref{eq:Wphi}) (solid line) for strips with metric from $\rho_1$, with
thickness $t/w=0.025$ and parameter $u_0/w=5/\pi$ for 2-, 3-, and
4-wrinkle strips (red (left), green (middle), and blue (right) curves, 
respectively). The $x$-axis is $\partial_u\rho=\cosh(u_0/R)$, so 
prescribed curvature increases to the
right. Error bars mark the variance of $\cos\phi$ on the center line
as $v$ runs from 0 to $2\pi$.
Beneath the plot is a sequence of 2-wrinkle strips showing the symmetry-breaking
transition that causes the conical strip ansatz to breaks down at
large curvatures.}
\end{figure}

Similar levels of agreement are obtained for other choices of metric,
curvature and thicknesses, though thinner sheets have a closer
agreement.  We tested 2-, 3-, and 4-wrinkle strips of thickness
$t/w=0.025$ with metric from
$\rho_1$ (with $u_0/w=5/\pi$ as are all strips with $\rho_1$ in this
paper) starting from
$R=150$ and increasing the curvature down to $R\approx 20$.  Fig.
\ref{fig:anglecompare} shows the resulting average $\cos\phi$ on the center
line as a function of $\partial_u\rho=\cosh(u_0/R)$, where the error bars show the
variance of $\cos\phi$.  Our theory and Eq. (\ref{eq:Wphi}) predict a linear
relationship, with zero variance.  This is observed up to fairly high
curvatures -- at the curvatures where $\cos\phi$ begins to oscillate,
we observe an intriguing symmetry-breaking from strips with $D_{md}$ 
symmetry to strips with $C_{mv}$ symmetry, and the wrinkles curl
into tubes.  The explanation of this transition is
outside the scope of this paper, but we display this transition in
2-wrinkle strips in the inset of Fig. \ref{fig:anglecompare}.  We note
that in this symmetry-breaking transition, as in the minimization
without a bias towards a set number of wrinkles, the system seems to
take a long time to find a minimum, often oscillating between
different states.

\begin{figure}
\includegraphics[width=3.25in]{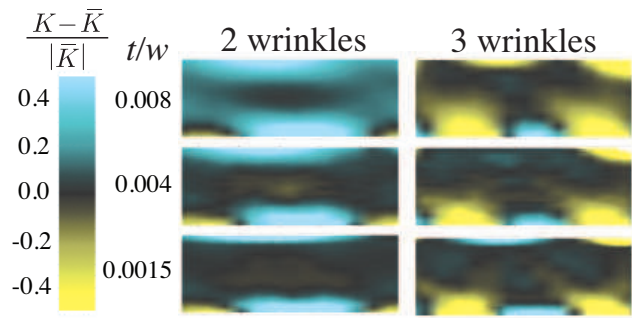}
\caption{\label{fig:Kseries} (color online) Gaussian curvature for
strips with $t/w \approx 0.008$, $0.004$ and $0.0015$ for $m=2$ and
$m=3$. These strips have a prescribed metric of $\rho_1$ with $\bar{K}
w^2 = -0.0324$. Only the middle third of the strip width is shown.
Darker regions have Gaussian curvature nearer to the prescribed curvature 
and so have less strain, while cyan or yellow (lighter) regions have 
more strain.}
\end{figure}

\begin{figure}
\includegraphics[width=3.25in]{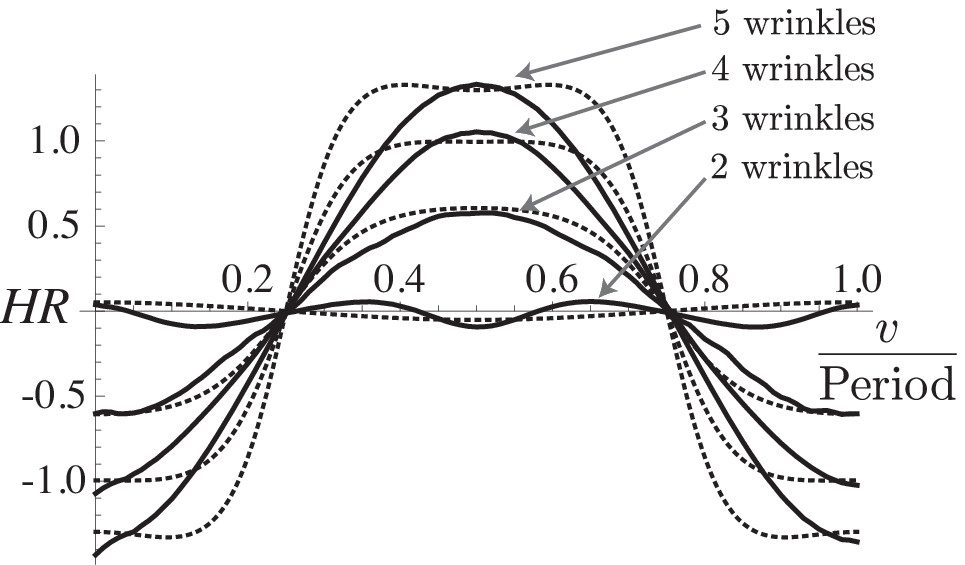}
\caption{\label{fig:H} The mean curvature $H R$ along the center line
of a $\rho_1$ strip with $t/w = 0.006$ and $\bar{K} w^2 = -0.029$ with
$5$, $4$, $3$, and $2$ wrinkles (solid) compared to theoretical
predictions (dotted) along the center line.}
\end{figure}

In Fig. \ref{fig:Kseries}, the distribution of Gaussian curvature is plotted for three thicknesses ($t/w \approx 0.008$, $0.004$ and $0.0015$) and $\bar{K} w^2 = -0.029$ for prescribed metric $\rho_1$. A boundary layer is evident in all the images. In both cases the degree of stretching
decreases near the center of the strip as the thickness decreases.

\begin{figure}
\includegraphics[width=3.25in]{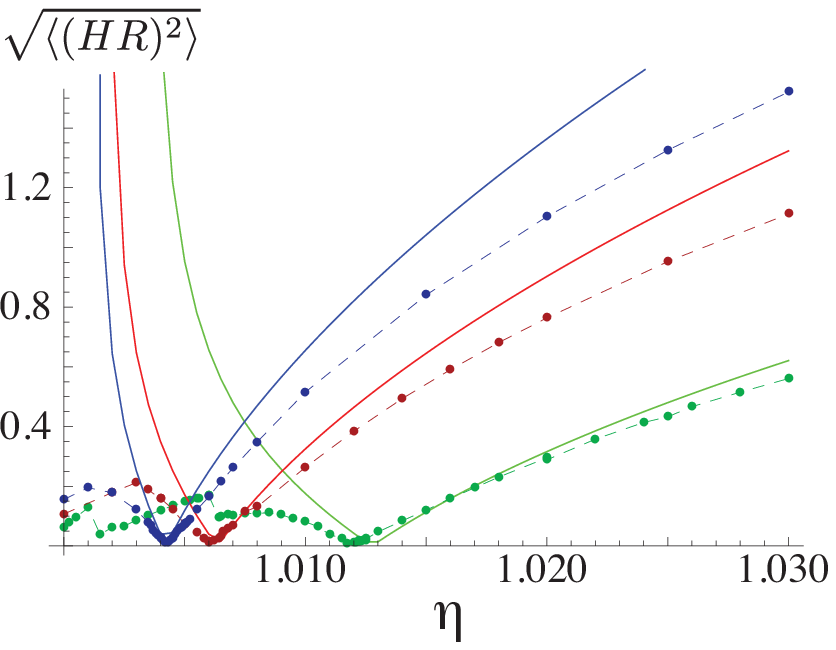}
\caption{\label{fig:resonancecompare} (color online) Comparison of
numerical mean curvature variance $\sqrt{\langle(HR)^2\rangle}$
(dashed lines are meant to guide the eye between the data points) to 
analytical predictions (solid) for $\rho_2$, $t/w=0.004$ as a function of $\eta$ for 2-,
3- and 4-wrinkle strips (green, red and blue, respectively) -- note
the observed resonance minima at roughly $\eta=1.0120$, $\eta=1.0060$ and
$\eta=1.0042$, compared to analytical predictions of $\eta=1.0128$,
$\eta=1.0063$ and $\eta=1.0042$, for 2-, 3- and 4-wrinkle strips,
respectively.}
\end{figure}

\begin{figure}
\includegraphics[width=3.25in]{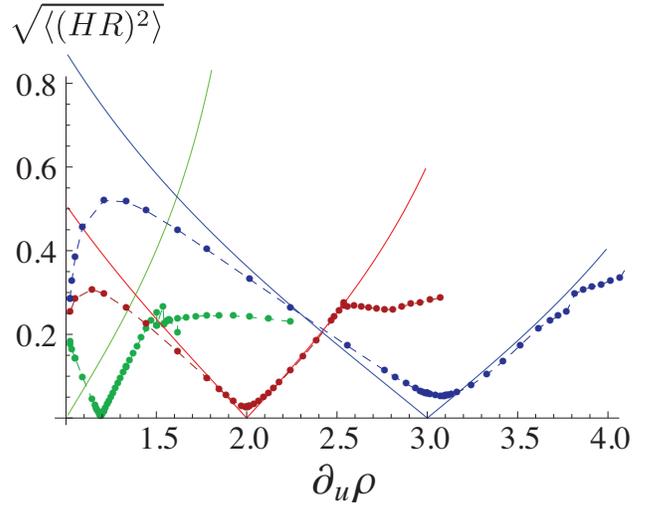}
\caption{\label{fig:resonancecomparemetric1} (color online) Comparison of
numerical mean curvature variance $\sqrt{\langle(HR)^2\rangle}$
(dashed lines are meant to guide the eye between the data points) to 
analytical predictions (solid) for $\rho_1$, $t/w=0.025$ as a function
of $\partial_u\rho=\cosh(u_0/R)$ for 2-,
3- and 4-wrinkle strips (green, red and blue, respectively) -- note
the observed resonance minima at roughly $\partial_u\rho=1.193$,
$\partial_u\rho=1.997$ and $\partial_u\rho=3.073$, compared to analytical
predictions of $\partial_u\rho=1$, 2 and 3, for 2-, 3- and 4-wrinkle strips,
respectively.}
\end{figure}

\begin{figure*}
\includegraphics[width=7in]{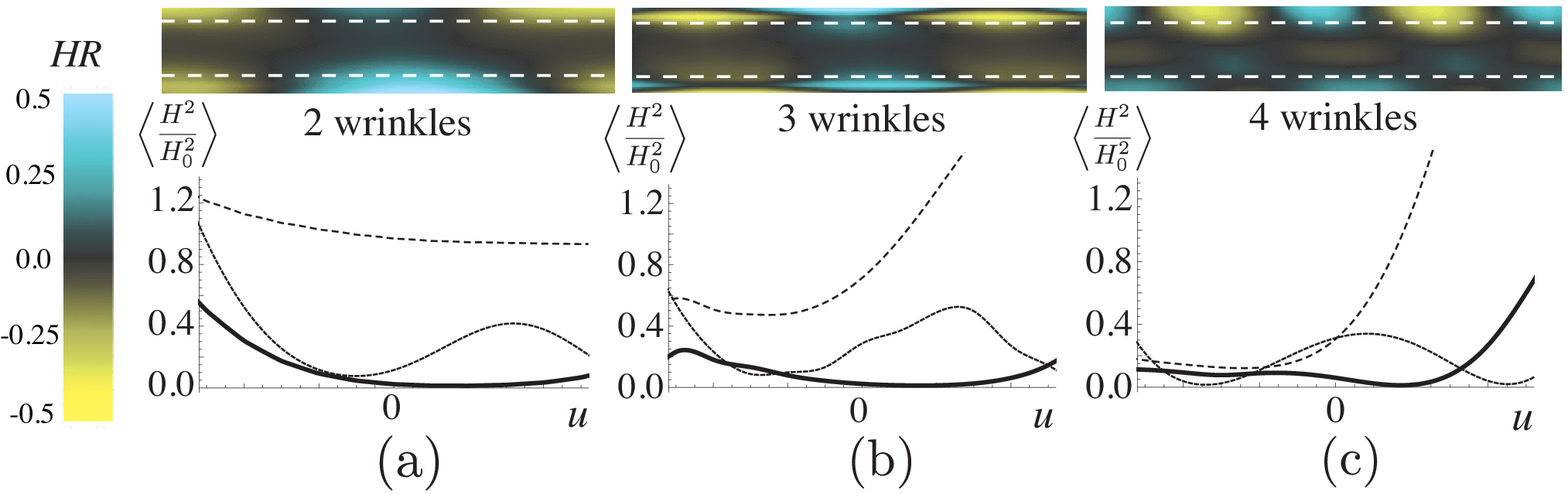}
\caption{\label{fig:resonance} (color online) Plots of $\langle
H^2/H_0^2 \rangle = \int dv~\rho(u)~[H^2(u,v)/H_0^2]/[2 \pi \rho(u)]$
for strips of width $w$ in the region $-w/4 < u < w/4$, where $H_0$ is the
average mean curvature within this region. The prescribed metric is
$\rho_2(u)$ and the thickness is $t/w=0.00375$. (a) $\langle H^2/H_0^2 \rangle$ for $\eta=1.02$ (dashed),
$\eta=1.01$ (solid) and $\eta = 1.00634$ (dotted) for $m=2$. The
resonance condition yields $\eta \approx 1.01275$. (b) $\langle
H^2/H_0^2 \rangle$ for $\eta=1.01$ (dashed), $\eta=1.00634$ (solid)
and $\eta = 1.0042$ (dotted) for $m=3$. The resonance condition yields
$\eta \approx 1.00634$. (c) $\langle H^2/H_0^2 \rangle$ for
$\eta=1.00634$ (dashed), $\eta=1.0042$ (solid) and $\eta = 1.003$
(dotted) for $m=4$. The resonance condition yields $\eta \approx
1.0042$. Above is the mean curvature of each strip
near resonance with dashed lines indicating the limits of the plots
below; the mean curvature is clearly extinguished (dark regions) 
near the center of the strip but not near the edges. }
\end{figure*}

We further test our model by comparing the numerically obtained $H$
along the center line with our analytical calculations. In Fig.
\ref{fig:H}, we show the specific case with prescribed metric
$\rho_1$, $t/w = 0.006$ and $\bar{K} = -0.029$. The solid lines show
$H/\sqrt{|\bar{K}|}$ along the center line for strips with different
numbers of wrinkles from $m=2$ to $m=5$. For the purposes of
comparison, we rescale $v$ to contain one period of the strip, $2
\pi/m$. Though these curves agree favorably in magnitude with those
predicted by our theory, there is some discrepancy in the precise
strip shape. This discrepancy may be caused by the failure of the
Gaussian curvature to agree with the prescribed curvature exactly or
perhaps by the fact that the analytical center line may approximate a
curve on the strip that is not exactly in the center.

A more stringent test of the analytical theory is the surprising
resonance condition of Eq. (\ref{eq:minimalbeta}). Since resonances
are at relatively high curvatures it is more difficult to achieve low
thicknesses without creases using metric $\rho_1(u)$, we focus on 
the metric $\rho_2(u)$ with various values of $\eta$. We choose $t/w =
0.00375$ and $\bar{K} w^2 = -0.01$; though very thin, even at this
$t/w$ there is some deviation of $K$ along the center line from its
prescribed value.  In Fig. \ref{fig:resonancecompare}, we show the
numerically measured variance of mean curvature and the analytical
predictions from the ansatz in Sec. \ref{sec:ansatz} for such strips.  It is clear
that the shape of the variance as a function of $\eta$ is fairly
well-mimicked with $\eta$ above resonance, but there is less good
agreement below resonance.  In fact, the numerically minimized shapes
tend to look nearly flat at $\eta=1$, as opposed to the ansatz which 
predicts a singular strip.  For completeness we include Fig.
\ref{fig:resonancecomparemetric1} which shows this data for
the $\rho_1$ strips shown in Fig. \ref{fig:anglecompare} -- the
agreement appears to be worse as a result of the thickness being
around $t/w=0.025$.  It does appear for 3-wrinkle ribbons that the 
point at which the numerical data begins to differ from the predictions occurs near 
the same values of $\partial_u\rho$ where the symmetry-breaking 
transition causes the conical strip ansatz to fail.

For 2-wrinkle strips, at $\eta\approx1.0055$ and $\eta\approx 1.0010$
below resonance we observe fairly sharp and reproducible jumps in the 
variance of the mean curvature -- we do not currently have a good
explanation of this, as the strip shapes do not appear to change
dramatically over these values of $\eta$.

In Fig. \ref{fig:resonance}, we plot $\langle H^2(u)/H_0^2 \rangle = \int dv~\rho(u)
[H^2(u,v)/H_0^2]/[2 \pi \rho(u)]$ over the surface of numerically minimized strips
with two, three, and four wrinkles above, near, and below resonance.
Here $H_0$ is the average mean curvature in a certain region of the
strip.  There is a clear trend toward
extinguishing the mean curvature near the center line of the strip
when $\eta$ is tuned near the resonance condition of Eq.
(\ref{eq:minimalbeta}).

Eq. (\ref{eq:minimalbeta}) and its application to $\rho_2$ strips, Eq.
(\ref{eq:respred}), predicts rather well the observed location
of the resonance. We find that the center line of a two wrinkle strip
will have $H=0$ when $\eta \approx 1.01275$, a three wrinkle strip will
have $H=0$ when $\eta \approx 1.00634$, and a four wrinkle strip will
have $H=0$ when $\eta \approx 1.00422$. 

For example, with two wrinkle strips we find that the mean curvature is
slightly smaller with $\eta = 1.012$ than near the predicted resonance of $\eta
= 1.013$. We ascribe this small discrepancy to our assumption that the
Gaussian curvature along the center line agrees with that prescribed
by the metric. Indeed, thick membranes do not show any evidence of the
resonance condition at all, nor does the Gaussian curvature reach its prescribed curvature on the center line. 
It is also interesting to note that, below resonance, the mean
curvature near the center line of a strip is $\pi$ out of phase with
the mean curvature near the strip edges. 


\begin{figure}
\includegraphics[width=3.25in]{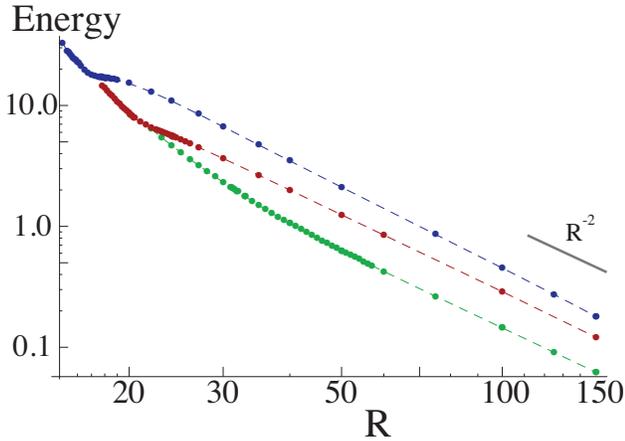}
\caption{\label{fig:metric1energy} (color online) Energy for $\bar{\rho}=\rho_1$ as a function of $R = 1/\sqrt{-\bar{K}}$ for 4-,3- and 2-wrinkle strips (from top to bottom; blue, red and green respectively).}
\end{figure}
\begin{figure}
\includegraphics[width=3.25in]{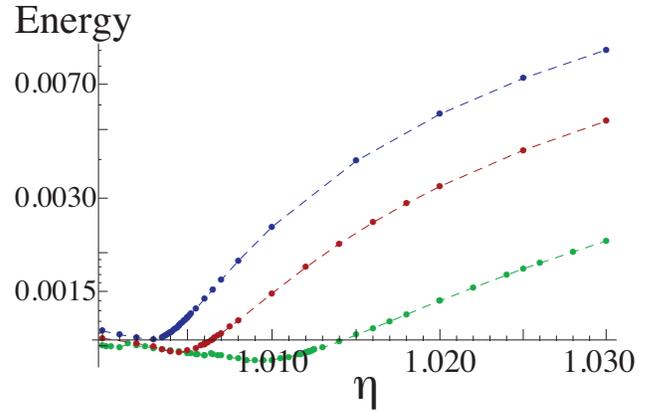}
\caption{\label{fig:metric2energy} (color online) Energy for $\bar{\rho}=\rho_2$ as a function of $\eta$ for 4-,3- and 2-wrinkle strips (from top to bottom; blue, red and green respectively).}
\end{figure}

Finally, we plot the energy for $\rho_1$ and $\rho_2$ in Figs. \ref{fig:metric1energy} and \ref{fig:metric2energy}. In both cases, the energy of $m=2$ appears to be lowest throughout the range of parameters tested, though for a large degree of swelling the energies become nearly degenerate. For Fig. \ref{fig:metric2energy}, dips in the energy occur at the resonances in $\eta$ as we might expect, since the dominant term in the bending energy vanishes near the resonances. In Fig. \ref{fig:metric1energy}, the resonances occur in the regions where $R$ is small and the energies are nearly degenerate.  At large $R$, the energy scales as $1/R^2 = |\bar{K}|$, which we expect since $E_B \sim H^2 \propto 1/R^2$.

Note that, at a resonance, the energy is dominated by higher order terms that we have neglected in our analysis. It is these additional terms that set the energy scale at the resonances. It is these higher order terms that seem to determine which shapes have the lowest absolute energy.

\section{Conclusion}\label{sec:conclusion}

In this paper, we have studied the interplay between topology and buckling for elastic strips which have been swelled inhomogeneously. We have focused on negative Gaussian curvatures with metrics that depend only on distance from the strip center line, though nothing precludes studying more general metrics. We identify a class of closed strips for which we can guarantee a prescribed Gaussian curvature; these ``conical'' strips have the property that their unit normals point along a cone everywhere on their center line. Using this family of strips, we predict the existence of minimal bending resonances for which the mean curvature along the strip center line vanishes for a prescribed number of wrinkles. This surprising result was confirmed in numerical simulations. To our knowledge, this phenomenon has not been observed experimentally in any system as of yet.

The main surprise from our numerical tests was the fact that conical
strips were such a good description of the numerical strip shape. It
is not clear if this agreement can persist to very thin strips, nor is
it clear if there are other, lower energy, strip shapes which do not
exhibit this property. Strictly speaking, our analytical calculations
are only valid within this class of strips and, at best, our resonance
condition is sufficient, but not necessary, to find a strip with $H=0$
along the center line. Further work quantifying closed strip shapes
will be necessary to understand this fortuitous agreement.  For
instance, there may exist an expansion of the elastic energy
around conical solutions which should explain the deviation from
conical strips at very high curvatures.


Though the experiments of Klein \textit{et al.} \cite{klein07} are for
wide strips with $w \sqrt{-K} \sim 1$, our results do shed some light
on the frustration that may be leading to the high degree of wrinkling
observed. Unlike experiments \cite{sharon-priv}, our numerical
minimization shows no indication that the number of wrinkles increases
with decreasing thickness.  Whether this is due to the narrowness of our
ribbons compared to the experiments is not clear. Indeed, it would be interesting to probe
the degree to which strips with various numbers of wrinkles are
metastable by perturbing the minimized strips and re-minimizing. 

Another interesting point about our analytical predictions is
that only the values at $u=0$ of the first two derivatives of $\rho$ with
respect to $u$ enter the formulas.  Preliminary tests of minimizing
strips with an added $u^3$ term to $\rho_1$ indicated that the
behavior of the Gaussian curvature on the center line changed by a
bit, but the angle of the normal was the same and the mean curvature was also close,
so we expect that more systematic tests will confirm our predictions 
in their regime of validity.

It is also interesting to consider the results of the numerical
simulations more closely. One of the generic features of all the
numerical minima is the persistence of strain in the strip, in
particular, in fairly localized regions of the wrinkle. On the
boundaries this is plausible as one might expect a boundary layer to
appear on quite general grounds. On the other hand, even along the
center line there is a small degree of strain, and we are unable to
obtain strips that are sufficiently thin for this strain to vanish.
Though we do not do so here, this residual strain can be studied using
a more careful perturbative expansion of the energy \cite{chen}.

\begin{acknowledgements}
We acknowledge useful discussions with E. Efrati,
E. Sharon, and R. Kupferman. We are particularly indebted to them for providing preprints of their recent work. CDS acknowledges funding from the National Science
Foundation Award No. DMR-0846582. BGC acknowledges the hospitality of the
UMass Soft Matter Summer School, where this work was
begun and the encouragement of R. Kamien as well as support from NSF
Grants No.
DMR05-47230 and DMR05-20020.
\end{acknowledgements}

\appendix

\section{Conical, closed strips}\label{sec:appendixB}

Recall that we wish to find closed solutions of 
$\partial_v\mathbf{v_x}(v)=\mathbf{B}(v)\times\mathbf{v_x}(v)$, where 
\begin{equation} 
\textbf{B}(v) = -\frac{h_{v v}}{\rho} \hat{i} + h_{u v} \hat{j} -
\partial_u\rho \hat{k}.
\end{equation}

In this form, the nontrivial dependence of $\textbf{B}$ on $v$ makes
it difficult to determine the strip shape or to determine generally
sufficient and necessary conditions for the strip to be closed.  Of
course, $\textbf{B}$ must be $2\pi$-periodic and we shall assume this
in what follows.  Note furthermore that this linear, homogeneous equation 
may be solved by Fourier analysis, though the conditions for
closedness lead to an infinite set of equations relating the Fourier
coefficients of $\rho$ and $h_{ij}$.

To motivate our construction, consider the special case of $\textbf{B}$ 
with a constant direction but possibly changing magnitude.
Closed solutions for $\mathbf{v_x}$ to this problem rotate in a plane 
perpendicular to $\textbf{B}$, with a frequency that is proportional
to the magnitude of $\textbf{B}$.  If we rotate $\textbf{B}$ so that
it's pointing in the $k$-axis, then:
\begin{align*}
\mathbf{v_x}(v)&=|v_x|[\cos(\omega v)\hat{i}+\sin(\omega v)\hat{j}]\\
\omega&=|B|
\end{align*}

The class of solutions that we consider build off of this case; 
we allow the direction of $\textbf{B}$ to move on a cone.  If we allow
a rotation of the $ijk$ coordinate frame by a $v$-dependent matrix, 
$\textbf{O}(v)$, then $\textbf{B}$ may be transformed into a vector
that is always parallel to the $k$-axis.  
We will consider the
solutions in this class for which the direction of $\textbf{B}$ lies
along a cone with axis the $k$-axis.  

To make a bit more sense of this, remember that the $k$-axis
corresponds to components of the normal vector -- as it turns out,
this restriction will force the normal vector to point along a cone
oriented along the $z$-axis in space.  This is why we will call these
conical strips.

Due to this specialization, we are restricting the allowed values of
the second fundamental form to a subclass of possible strip shapes.
Our form for $\mathbf{B}$ remains general enough, however, that many
solutions for closed strips still exist in the set of conical
strips, including some with low bending energies.  Furthermore,
the center lines of strips we find by numerical minimization 
have this property as well.

We proceed by rotating $\mathbf{B}$ about the $k$ axis with respect to
an arbitrary angle $-\beta(v)$.
This yields a modified equation $\partial_v \mathbf{v_x}' =
\textbf{B}' \times \mathbf{v_x}'$, where
\begin{eqnarray}
\textbf{B}' &=& -\left(\frac{h_{v v}}{\rho} \cos \beta + h_{v u} \sin \beta\right) \hat{i}\\
& & - \left(-h_{v u} \cos \beta + \frac{h_{v v}}{\rho} \sin \beta\right) \hat{j}\nonumber\\
& & + \left(\partial_v \beta-\partial_u \rho\right) \hat{k}.\nonumber
\end{eqnarray}
We now choose $\beta$ so that the $\hat{i}$ component of $\mathbf{B}'$
vanishes, which simplifies the equation somewhat. In particular, we
choose
\begin{align}
\frac{h_{vv}}{\rho}\cos\beta&=-h_{vu}\sin\beta\nonumber\\
\rho H-\rho \sqrt{H^2-K}\sin\alpha\cos\beta&=
-\rho \sqrt{H^2-K}\cos\alpha\sin\beta\nonumber\\
(\sin\xi-\sin\alpha)\cos\beta&=-\sin\beta\cos\alpha\nonumber\\
\tan \beta&=\frac{\sin\alpha-\sin\xi}{\cos\alpha},
\end{align}
where $\sin \xi \equiv H/\sqrt{H^2-K}$ and we've used Eqs. (\ref{eq:ll}). 
The vector $\mathbf{B}'$ will
therefore point in a direction in the (new) $jk$-plane.

Let $\phi$ be the constant angle between $\mathbf{B}'$ and the 
$k$ axis, defined by
\begin{align}
\tan\phi&=\frac{\mathbf{B}'_j}{\mathbf{B}'_k}\nonumber\\
&=\frac{-h_{v u} \cos \beta + 
\frac{h_{v v}}{\rho} \sin \beta}{\partial_u \rho-\partial_v\beta}\nonumber\\
&=-\frac{\sqrt{\rho \partial_u^2 \rho}\cos\alpha}
{\cos\beta\cos\xi(\partial_u \rho-\partial_v\beta)}.
\end{align}
Note that we've applied the relation $K=-\partial_u^2 \rho/\rho$.

By eliminating $\alpha$, we derive that
\begin{align}\label{appeq:meancurvature}
\frac{H}{\sqrt{-K}} &= \tan \xi \\
 &= \frac{1}{2 \sin \beta} \left( \frac{\tan\phi(\partial_u\rho - 
\partial_v \beta)}{\sqrt{\rho \partial_u^2 \rho}}-\frac{\sqrt{\rho \partial_u^2\rho}}
{\tan\phi (\partial_u \rho -\partial_v \beta)}\right)\nonumber.
\end{align}

This expression for the mean curvature tells us that for conical strips,
$\phi=0$ (flat disk limit) or $\phi=\pi/2$ (cylinder limit) is 
potentially bad.

As above, a closed particle trajectory requires that the particle velocity,
$\mathbf{v_x}'$, will lie in the plane perpendicular to $\mathbf{B}'$.
The instantaneous frequency (in $v$) of this motion in the plane is given by
$\omega(v)=( \partial_v \beta- \partial_u \rho ) \sec\phi$.
To find the complete solution, we return to our original unrotated
frame by rotating the vector $\mathbf{v_x}'$ by $-\beta$ to find
\begin{widetext}
\begin{eqnarray}
\mathbf{v_x}(v) &=& \left[\sin\phi\cos \gamma \sin W-\cos\phi \sin \gamma 
\right] \hat{k}+\left[\cos \gamma \cos \beta \cos W+ \cos\phi\cos
\gamma \sin \beta \sin W + \sin\phi\sin \gamma\sin \beta\right] \hat{i}\nonumber\\\label{eq:vsoln}
& &+ \left[ - \cos \gamma \sin \beta \cos W +\cos\phi\cos \gamma \cos
\beta \sin W + \sin\phi\sin \gamma\cos \beta \right] \hat{j},
\end{eqnarray}
\end{widetext}
where $\gamma$ is an integration constant having to do with motion of
the fictitious particle along the $\mathbf{B}'$ axis and $W=\int_0^v
dv' \omega(v')+W_0$.  The constant $W_0$ sets the initial phase of the
motion. Thus $W$ is defined so that $W(0)=0$, or
$W=\sec\phi\int_0^vdv'(\partial_v\beta-\partial_u \rho)$.

Recall that the vector $\mathbf{v_x}$ contains the $\hat{x}$
components of the entire frame. But we can also use this solution for the
components $\mathbf{v_y}$ and $\mathbf{v_z}$ of the strip frame, each
of which is described by the solution given in equation
(\ref{eq:vsoln}) and differing only in the integration constants
$\gamma$ and $W(0)=W_0$. These constants must be chosen to be
compatible with the orthogonality of the three vectors describing the
frame. We choose: (1) $\gamma=0$ and $W_0=0$, (2) $\gamma=0$ and
$W_0=\pi/2$, and (3) $\gamma=-\pi/2$ to define a set of three mutually orthogonal vectors at $v=0$. Since the evolution acts by rotations, these three orthogonal vectors remain orthogonal throughout the evolution. 

If we take the $\hat{i}$ components of the three solutions we get
\begin{eqnarray}
\partial_u \textbf{r} &=& \sin^2 \frac{\phi}{2} \left[\cos(W+\beta) \hat{x} - \sin(W+\beta) \hat{y} \right] \nonumber\\
& & + \cos^2 \frac{\phi}{2} \left[\cos (W-\beta) \hat{x} - \sin (W-\beta) \hat{y} \right]\nonumber\\
& & - \sin \beta \sin \phi \hat{z}.
\end{eqnarray}
If we take the $\hat{j}$ components of the three solutions, we obtain
\begin{eqnarray}
\frac{\partial_v \textbf{r}}{\rho} &=& - \sin^2\frac{\phi}{2} \left[\sin (W+\beta) \hat{x} + \cos(W+\beta) \hat{y}\right]\nonumber\\
& & + \cos^2\frac{\phi}{2}\left[ \sin(W-\beta) \hat{x} + \cos(W-\beta) \hat{y} \right]\nonumber\\
& & - \cos \beta\sin\phi \hat{z},
\end{eqnarray}
Finally, the $\hat{k}$ components yield the normal vector
\begin{eqnarray}
\norm = \sin\phi\sin W\hat{x} + \sin\phi\cos W \hat{y} + \cos\phi\hat{z}.
\end{eqnarray}


\begin{thebibliography}{99}

\bibitem{drasdo} D. Drasdo, Phys. Rev. Lett. \textbf{84}, 4244 (2000).

\bibitem{mahadevan09} H. Liang and L. Mahadevan, Proc. Nat. Acad. Sci. \textbf{106}, 22049 (2009).

\bibitem{marder03a} M. Marder, Found. Physics \textbf{33}, 1743 (2003).

\bibitem{sharon07} E. Sharon, B. Roman and H.L. Swinney, Phys. Rev. E \textbf{75}, 046211 (2007).

\bibitem{nath03} U. Nath, B.C.W. Crawford, R. Carpenter and E. Coen, Science \textbf{299}, 1404 (2003).

\bibitem{benamar05} A. Goriely and M. Ben Amar, Phys. Rev. Lett. \textbf{94}, 198103 (2005).

\bibitem{swinney02}
E. Sharon, B. Roman, M. Marder, G.S. Shin and H.L. Swinney, Nature (London) \textbf{419}, 579 (2002).

\bibitem{mora06} T. Mora and A. Boudaoud, Eur. Phys. J. E \textbf{20}, 119 (2006).

\bibitem{klein07} Y. Klein, E. Efrati, and E. Sharon, Science \textbf{315}, 1116 (2007).

\bibitem{landau} L.D. Landau and E.M. Lifschitz, \textit{Theory of Elasticity, 3rd Ed.} (Pergamon, London, 1959).

\bibitem{benamar08} J. Dervaux and M. Ben Amar, Phys. Rev. Lett. \textbf{101}, 068101 (2008).

\bibitem{nechaev} S. Nechaev and R. Voituriez, J. Phys. A \textbf{34}, 11069 (2001).

\bibitem{audoly02}
B. Audoly and A. Boudaoud, C.R. Mecanique \textbf{330}, 831 (2002).

\bibitem{audoly03}
B. Audoly and A. Boudaoud, Phys. Rev. Lett. \textbf{91}, 086105 (2003).

\bibitem{marder03} M. Marder, E. Sharon, S. Smith, and B. Roman, Europhys. Lett. \textbf{62}, 498 (2003).

\bibitem{marder06} M. Marder and N. Papaniolaou, J. Stat. Phys. \textbf{125}, 1065 (2006).

\bibitem{efrati08a} E. Efrati, E. Sharon, R. Kupferman, J. Mech. Phys. Sol. \textbf{57}, 762 (2009).

\bibitem{santangelo09} C.D. Santangelo, EPL \textbf{86}, 34003 (2009).

\bibitem{embeddingbook}
Q. Han and J-X. Hong, \textit{Isometric Embedding of Riemannian Manifolds in Euclidean Spaces}, \textit{Mathematical Surveys and Monographs, vol. 130} (American Mathematical Society, Providence, RI , 2006)

\bibitem{poznyak} E.G. Poznyak and E.V. Shikin, J. Math. Sci. \textbf{74}, 1078 (1995).

\bibitem{arizona} J. Gemmer and S. Venkataramani, e-print arXiv:1005.4442

\bibitem{chen} B.G. Chen and C.D. Santangelo, (unpublished).

\bibitem{ciarlet} P.G. Ciarlet, J. Elasticity \textbf{78-79}, 1 (2005).

\bibitem{DoCarmo-book} M.P. Do Carmo, \emph{Differential Geometry of Curves and Surfaces},  (Prentice-Hall, Englewood Cliffs, NJ, 1976).

\bibitem{weingarten} D.J. Struick, \textit{Lectures on Classical Differential Geometry}, 2nd ed (Addison-Wesley, Reading, Mass, 1961).

\bibitem{kupferman} E. Efrati, E. Sharon and R. Kupferman (private communication).

\bibitem{sadowsky} M. Sadowsky, Sitzungsber. Preuss. Akad. Wiss. \textbf{22}, 412 (1930).

\bibitem{physicstoday} M. Marder, R.D. Deegan and E. Sharon,
Phys.Today \textbf{60}(2), 33 (2007).





\bibitem{sharon-priv} E. Sharon, private communication (2010).




\end{thebibliography}
\end{document}